# Wideband Bandpass Filters Based on Multipole Resonances of Spoof Localized Surface Plasmons

Xuanru Zhang,[1,2] Di Bao,[1,2] Jun Feng Liu,[1,2] and Tie Jun Cui[1,2,a]

[1] State Key Laboratory of Millimeter Waves, Southeast University, Nanjing 210096, China.

[2] Synergetic Innovation Center of Wireless Communication Technology, Southeast University, Nanjing 210096, China.

[a] Author to whom correspondence should be addressed. Email: tjcui@seu.edu.cn

**Abstract**

We propose wideband bandpass filters based on multipole resonances of spoof localized surface plasmons (SLSPs). The resonance characteristics and geometric tunability of SLSPs are investigated under microstrip excitations. Strong coupling with interlayer microstrip lines is proposed to join discrete multipole resonances into a continuous and flat passband. The SLSP filters exhibit wide passbands in compact sizes and well-balanced shapes, while holding satisfactory spurious rejection bands, group delays, and geometric tunability. This work exposes the SLSPs' application potential in filters as novel resonators.

Filters play essential roles in microwave systems, such as wireless communications, radars, *etc*. Planar microstrip filters have been widely used for the advantages of low cost, easy fabrication, easy integration with microstrip circuits, small sizes, and so on. Formulaic synthesis methods are taken for filters with relatively narrow passband (usually<20%), and have been realized in structures of coupled line, coupled resonators, *etc*.[1, 2] For wideband filters, multimode resonators are usually adopted, such as patch resonators, ring resonators, and stepped-impedance resonators.[3-5] The passband bandwidths of patch and ring resonators are still limited, though various perturbations are introduced to split up the degenerate modes.[1, 6] Ultrawide bandwidth as high as 100% can be achieved in stepped-impedance resonators.[4, 5] However, the stepped-impedance resonators are always length consuming, thus prevent them from applications in ultra-compacted circuits.[3]

Spoof localized surface plasmons (SLSPs) refer to periodically textured metallic cylinders (2D) or ultrathin metallic disks (3D), which resemble multipole resonances of optical localized surface plasmons (LSPs) in microwave and Terahertz spectra.[7] SLSPs exhibit conceptual similarity to the propagating mode of spoof surface plasmon waves (SSPWs).[8-10] SSPWs have been widely studied in the past decade, and have found plenty applications as novel transmission lines in passive and active circuits.[11-15] Different from propagating SSPWs, SLSPs are localized resonant modes. Compared with conventional resonators such as stepped-impedance resonators, split ring resonators, etc., SLSPs possess more abundant resonant modes (both axial and radial multiple modes), thus, more complex physical phenomena and mechanisms can be expected.[16-23] SLSPs also exhibit the potential in compacting resonator sizes, stemming from their strong near-field enhancement. However, SLSPs have been theoretically proposed and investigated very recently (since 2012), and practical circuit devices have seldom been realized based on SLSPs (except one demonstration of SLSP sensor [24]).

In this paper, the multipole resonances and geometric tunability of SLSP disks are investigated, and utilized to construct microstrip wideband bandpass filters. Strong coupling with the interlayer microstrip lines is proposed, to join the multipole resonant peaks of the SLSPs into a flat passband. Two SLSP filters centered at 4.3 GHz and 11.7 GHz are validated, respectively. The SLSP filters exhibit wide passbands in compact sizes and balanced shapes, while holding satisfactory spurious rejection bands, group delays, and geometric tunability.

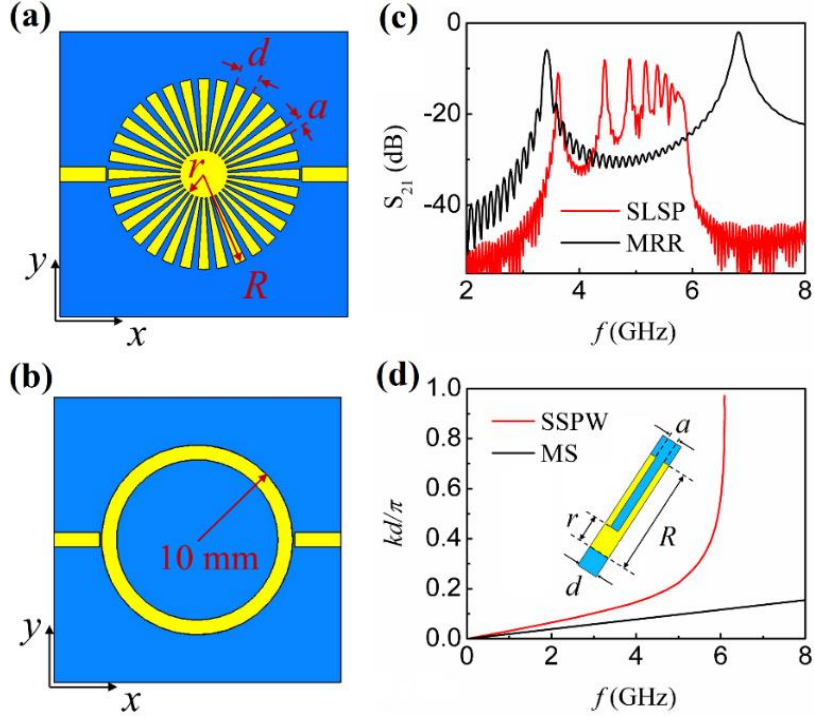

FIG. 1. (a, b) Schematics of the SLSP disk (a) and microstrip ring resonator (b). (c) Transmission spectra ($S_{21}$) of the microstrip excited structures in (a) and (b). (d) The dispersion curves of the corresponding SSPW and 50 Ω microstrip (MS) line. The inset shows a single unit of SSPW.

The multipole resonances of SLSPs have been investigated thoroughly, under the plane-wave excitation or monopole probe excitation. We simulate SLSP resonances under colinear microstrip excitation, and compare them with the microstrip ring resonator (MRR), as demonstrated in FIG. 1. All simulations in this paper are implemented in CST Microwave Studio. The SLSP disk is corrugated with N periodically radical grooves, with an outer radius of $R$ and an inner radius of $r$. Here, N is chosen to be 32, $R$ is 10 mm and $r$ =2.5 mm. The groove width $a$ equals 0.4$d$.

Thus, $a$ is 0.78 mm and $d$ is 1.96 mm. The whole structure is based on a Rogers RT 5880 substrate of 0.508 mm thick, with a large-area ground plane to enhance the resonances and integrate with the microstrip circuit.[25] The contrastive MRR is has the same outer radius of 10 mm, and a ring width of 1.54 mm (50 Ω). Both the two resonators are excited by colinear microstrip lines of 50 Ω, with a gap of 0.2 mm separated from the resonators. All the structures investigated here are covered by a prepreg layer of 0.1 mm (permittivity 3.5 and loss tangent 0.004) and a Rogers RT 5880 layer of 0.508 mm on top. The two covering layers are included so that the spectra fit those in the following filters. As shown in FIG. 1 (c), the second resonant peak of the MRR of 6.8 GHz locates perfectly at the second harmonic of the first peak of 3.4 GHz. SLSP exhibits a multipole resonant band, composed of densely spaced multiple resonances. As revealed by the modal expansion technique, the SLSP resonances can be interpreted as standing waves of SSPW around the disk perimeter.[7] The corresponding SSPW unit is shown in the inset of FIG. 1 (d), and the dispersion curves are rotated for correspondence to the spectrum in FIG. 1 (c). The nonlinear dispersion curve of the SSPWs compress the standing waves of different orders into a densely-spaced band. The edge of the SLSP resonant band corresponds to the cutoff band of the SSPWs.

The multipole resonances of SLSP are further verified by near-field distributions shown in FIG. 2, in which the red and blue colors indicate the positive and negative values respectively. Note that the color bars have been adjusted individually to show all the mode patterns clearly, so the relative strength of each mode cannot be identified from these field distributions. The slight asymmetry of the field distributions results from the asymmetric excitation (from left to right). These field distributions resemble those in plane-wave excitation and monopole probe excitation in the references,[7, 16, 17] and confirm that the microstrip line is an effective way to excite the resonant modes of SLSPs.

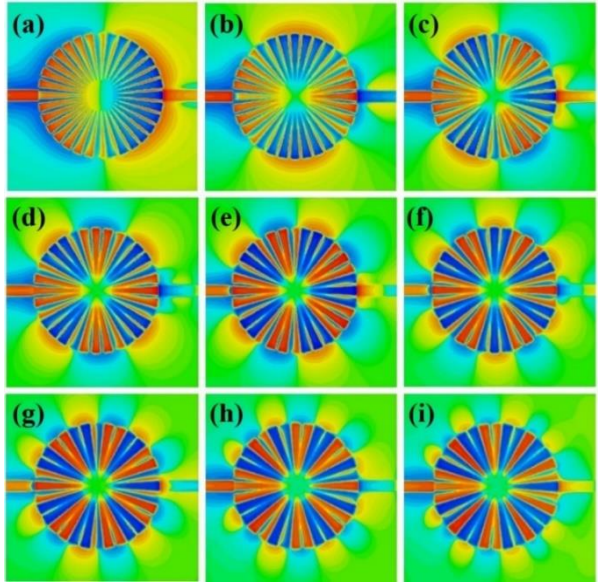

FIG. 2. Simulated $E_z$ distributions of the SLSP disk in FIG. 1 at a $x$-$y$ plane 0.2 mm above the sample surface, at (a) 3.62 GHz, (b) 4.44 GHz, (c) 4.88, GHz, (d) 5.18 GHz, (e) 5.38 GHz, (f) 5.52 GHz, (g) 5.66 GHz, (h) 5.74 GHz, (i) 5.80 GHz,. These correspond to the multipole resonant peaks in FIG. 1 (c). A rainbow color bar is used, with the red color referring to positive $E_z$ and the blue color referring to negative $E_z$.

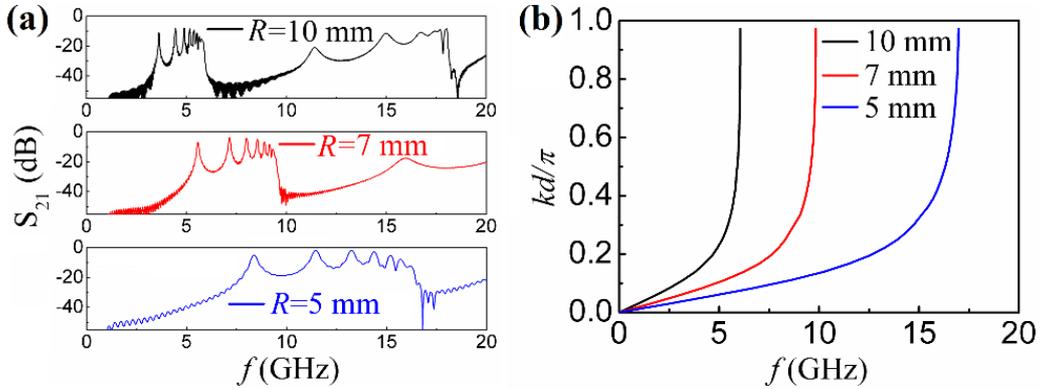

FIG. 3. Tunability of the SLSPs. (a) Transmission spectra of microstrip excited SLSPs, with different $R$ values and other geometric parameters fixed. (b) Dispersion curves of SSPWs corresponding to SLSPs of cases $R$=10 mm, 7 mm and 5 mm.

The geometric tunability of SLSPs and dispersion curves of the corresponding SSPWs are demonstrated in FIG. 3. The cases of $R$=10 mm, 7 mm and 5 mm are demonstrated, with other geometric parameters fixed. The cutoff frequencies increase with reducing $R$, leading to broadened distributions of the resonant peaks. These will result in different bandwidth in the following filter designs.

It is mentionable that the multipole peaks of the SLSPs are gradually denser at higher frequencies. These distributions are different from the Chebyshev poles, which are slightly sparser at higher frequencies.[3] However, the higher-order SLSP modes are more difficult to excite with lower $S_{21}$ magnitudes, as shown in FIG. 3(a). Denser peaks at higher frequencies can offset the difficulties in excitation, and lead to relatively flat spectra for the wideband filters.

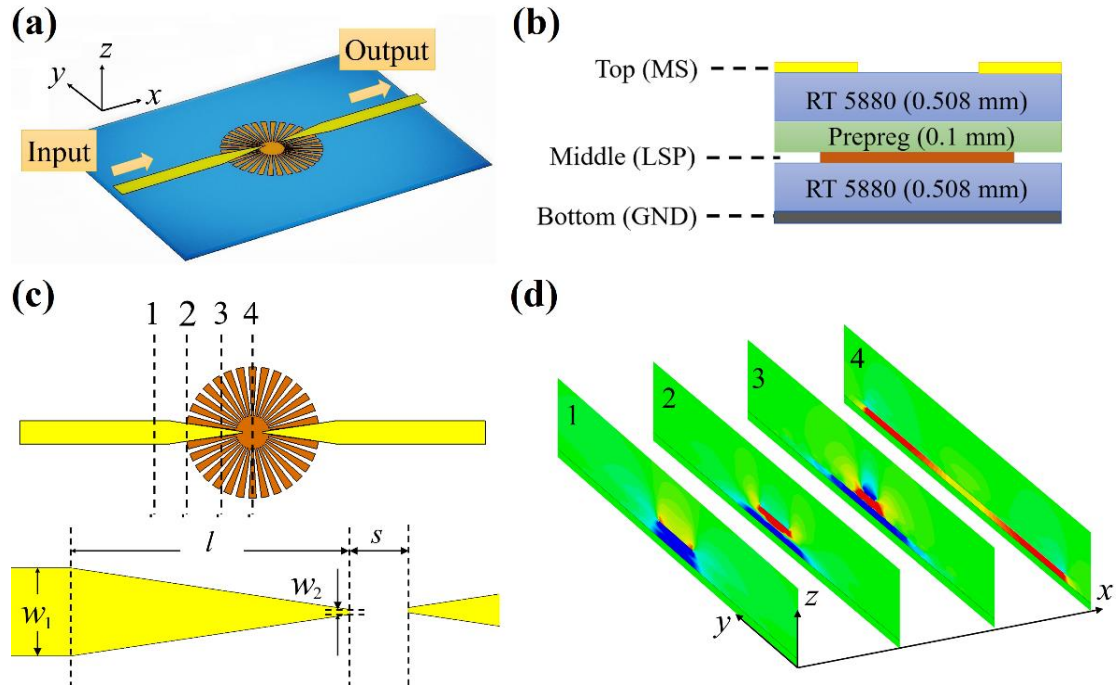

FIG. 4. (a) Schematic of the proposed SLSP filter. (b) Side view of the layered structure. (c) Detailed schematic of the microstrip excitation structure. (d) $E_z$ distributions at four perpendicular planes indicated in (c), at 3.52 GHz.

A microstrip interlayer coupling scheme is proposed as demonstrated in FIG. 4, to join all the discrete peaks into a flat passband. The whole structure is composed of double layers of Rogers RT 5880 substrate of 0.508 mm thickness, pasted together by a prepreg layer of 0.1 mm thickness. The SLSP disk is located at the middle layer, while the microstrip feed lines are located at the top layer. Copper layers are presented in different colors. The SLSP geometry is the same as that in FIG. 1 of $R$=10 mm. As demonstrated in FIG. 4 (c), the microstrip line is tapered off from $w_1$=3.5 mm width (50 Ω) to $w_2$=0.2 mm in an $l$ of 12.5 mm, to efficiently match the fine-structured field distributions of high-order modes. The two tapered tips are $s$=3 mm apart and are symmetric relative to the center. The gradual transition from the microstrip mode to the SLSP dipole mode at 3.52 GHz is demonstrated in FIG. 4 (d) for an intuitive view. The 1-4 planes are 15 mm, 10 mm, 5mm and 0 mm from the center. The dipole frequency here is slightly shifted from that of 3.62 GHz in FIG. 1 and 2, due to stronger coupling with the microstrip lines.

Stemming from the tunability of the SLSP resonances,[16] the SLSP filters are also geometrically tunable. We also design and fabricate a SLSP filter with $R$=5 mm. The tapered microstrip structure is changed to $l$=7.5 mm and $s$=2 mm for a better match for $R$=5 mm case. For the filter of $R$=10 mm, both the measured results and the simulated results in FIG. 5(a) exhibit a passband centered at 4.3 GHz and a 3-dB fractional bandwidth of 53%. The 3-dB fractional bandwidth is defined as the bandwidth of $S_{21}$> 3 dB divided by the center frequency. For the filter of $R$= 5mm, the simulated result exhibits a passband centered at 11.7 GHz and a 3-dB fractional bandwidth of 73%; while the measured result exhibits a deteriorative 3-dB fractional bandwidth of 65%. The measured $S_{21}$ degrades at high frequencies due to the limited working bandwidth of the SMA connector. The diameters of the SLSPs are 0.29 and 0.39 free-space wavelengths of the center frequencies respectively. Considering the effects of the substrate, the SLSPs diameters are 0.40 and 0.58 wavelengths of the center frequencies in the substrate. As all axial modes of the basic radial order are incorporated into the passband, the SLSP filters possess high out of band rejection. The spurious bands are due to radial high-order modes, which possess much sharper peaks and are difficult to be excited.[17] The first spurious harmonic occurs at 15 GHz (31 GHz), which is 3.4 (3.6) fold of the center frequency of the first passband, for $R$=10 mm ($R$=5 mm) case. The spurious bands are ignored in the $S_{21}$ curves in FIG. 5 to zoom in the passband.

7 / 12

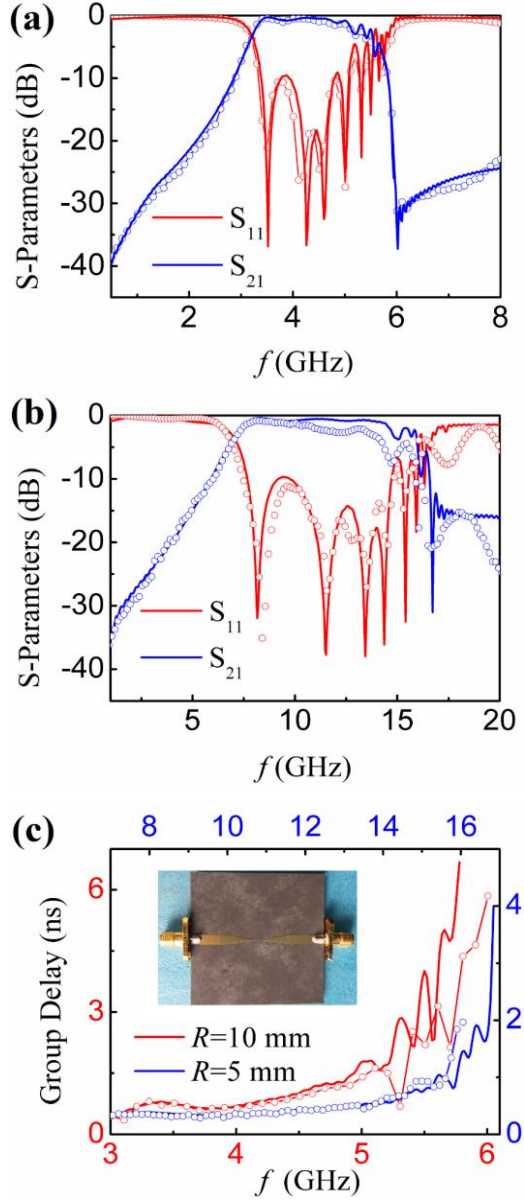

FIG. 5 (a, b) Simulated and measured S-parameters of the SLSP filter, with $R$=10 mm (a) and $R$=5 mm (b). The dotted lines denote the measured results. (c) Simulated and measured group delays of $S_{21}$. The inset is the photograph of the top layer of the fabricated filter of $R$=10 mm.

As demonstrated in FIG. 5(c), the group delays for both cases are around 1 ns at most of the passbands, while they increase rapidly around the edges of the passbands. This is due to the strong nonlinear dispersion of the corresponding SSPWs around the cutoff bands, as shown in FIG. 3. The sharp edges of the group delays are inevitable for all bandpass filters based on slow-wave structures.

TABLE I. Comparison with previous works.

| Reference | Center Frequency (GHz) | 3-dB Bandwidth (%) | Size ($\lambda r^2$) |
|---|---|---|---|
| [3] | 2.45 | 52.5 | 0.72 x 0.05 |
| [4] | 6.85 | 70 | 1.18 x 0.17 |
| [5] | 6.86 | 100 | 1.38 x 0.42 |
| This ($R$=10 mm) | 4.4 | 53 | 0.4 x 0.4 |
| This ($R$=5 mm) | 11.7 | 73 | 0.58 x 0.58 |

TABLE I compares this work with some typical previous works. As not all papers discuss on the spurious harmonics and group delays, we list only the common indices. All the sizes are normalized into the wavelengths at the center frequencies in the substrate, i.e., $\lambda_r = \lambda_0/\sqrt{\varepsilon_r}$. When more than one filters are validated in the papers, we choose the one of the best performance. As discussed in the introduction, though the stepped-impedance resonators can achieve a bandwidth as wide as 100% and the width of the circuit size can be highly compacted, they are inevitably length consuming. The comparison evidences the satisfactory bandwidth, balanced shapes and compactness of the SLSP filters.

The multipole resonances and geometric tunability of SLSPs are investigated. Wideband bandpass filters based on the SLSP disks are validated and compared with some typical previous works. The SLSP filters exhibit wide passbands in compact sizes and well-balanced shapes, while holding satisfactory spurious rejection bands, group delays and geometric tunability. This work exposes SLSPs' application potential in filters as novel resonators. Coupled SLSP resonators [22, 23] can be adopted in further research to introduce more physical phenomena and to construct devices with more outstanding performance.

This work was supported in part from the National Natural Science Foundation of China (61701108, 61735010, 61631007, 61571117, 61401089, 61522106), in part from the 111 project (111-2-05).